\documentclass[aps,pra,reprint,showpacs,twocolumn,superscriptaddress,eqsecnum]{revtex4-2}

\usepackage{graphicx}
\usepackage{dcolumn}
\usepackage{bm}
\usepackage{amsmath}
\usepackage{multirow}
\usepackage[utf8]{inputenc}
\usepackage{physics}

\usepackage{tabularx}
\usepackage{color}
\usepackage{appendix}
\usepackage{enumitem}
\usepackage{natbib,hyperref}
\usepackage{xr}
\hypersetup{colorlinks=true,linkcolor=blue,urlcolor=blue,citecolor=blue}

\newcommand{\RN}[1]{%
  \textup{\uppercase\expandafter{\romannumeral#1}}%
}

\begin{document}

\preprint{APS/123-QED}

\title{Hidden local adiabatic ramp in the modulated time evolution and the quantum approximate optimization algorithm}
\author{Zekun He}
 \email{zh168@georgetown.edu}
 \affiliation{%
Department of Physics, Georgetown University, Washington DC 20057, USA
}%

 \author{A.~F.~Kemper}
\email{akemper@ncsu.edu}
\affiliation{Department of Physics, North Carolina State University, Raleigh, North Carolina 27695, USA}

\author{J. K. Freericks}%
\email{james.freericks@georgetown.edu}
\affiliation{%
Department of Physics, Georgetown University, Washington DC 20057, USA
}%

\date{\today}

\begin{abstract}
Adiabatic state preparation provides an analytical solution for generating the ground state of a target Hamiltonian, starting from an easily prepared ground state of the initial Hamiltonian. While effective for time-dependent Hamiltonians with an energy gap to the first coupled excited state, the process becomes exceedingly slow as the gap narrows. Rather than strictly following the adiabatic theorem, a more robust approach allows controlled diabatic excitations during the evolution and numerically optimizes the path to eliminate these excitations by the end. In this work, this is achieved via modulated time evolution, using a time-dependent oscillating field $\lambda(t)$ to modulate the Hamiltonian, in conjunction with a transverse field $B(t)$ whose optimized shape closely resembles a local adiabatic ramp. Beyond modulated time evolution, the quantum approximate optimization algorithm (QAOA), which also employs a transverse field—defined as \( \beta(t)/\gamma(t) \)—exhibits a shape similar to the local adiabatic ramp. This resemblance offers a more intuitive and physically motivated way to understand the QAOA algorithm through the lens of time evolution.
\end{abstract}

\maketitle
\section{Introduction}

State preparation, especially ground state preparation, is a critically important algorithm for quantum computation. Adiabatic state preparation~\cite{born1928beweis,jansen2007bounds,farhi2000quantum} (or more practically, finite-time evolution using the local adiabatic ramp~\cite{roland2002quantum,richerme2013experimental}), is a simple way to create the ground state, but it requires an extremely long time evolution for high fidelity, which ultimately makes it impractical. Shortcuts to adiabaticity~\cite{guery2019shortcuts,hegade2021shortcuts,wan2004fast} can prepare the ground state in significantly shorter time frames, but they usually require evolving the system with additional complicated counter-diabatic Hamiltonians~\cite{demirplak2003adiabatic,demirplak2005assisted,berry2009transitionless,ieva2023Counterdiabatic}, which are prohibitive to implement. To improve the feasibility of these methods, various approaches have been explored, such as variational-based counter-diabatic Hamiltonians~\cite{ieva2023Counterdiabatic,li2024quantum}, optimal protocols in quantum annealing~\cite{brady2021optimal} and tensor network-based quantum circuit compression~\cite{Lubasch2024Towards}.

As we will show in this work, the key to accelerate adiabatic time evolution is to find a strategy that can return the amplitudes of diabatically excited states back to the ground state at the end of the time evolution, as shown in Fig.~\ref{fig:lamda B plot} panel (d). This is contrary to local adiabatic evolution, which is engineered solely to minimize the excitation out of the ground state, and does not optimize the possibility of returning amplitudes back to the ground state. One way to separately engineer a return mechanism is to incorporate a variational principle into the time evolution. We do so by introducing a time-dependent scaling field $\tilde{\lambda}(t)$, which modulates the scale of the Hamiltonian $\tilde{\lambda}(t)\hat{H}_0(t)$, and whose temporal profile is optimized to minimize the energy of the final evolved state.

In this work, we examine the ground state preparation of the long-range transverse-field Ising model. We denote the initial Hamiltonian by $\hat{H}_B$ (magnetic field term), and we write the (unmodulated) time-dependent Hamiltonian as $\hat{H}_0(t) = \hat{H}_A + B(t)\hat{H}_B$, such that the target Hamiltonian is $\hat{H}_0(t{=}t_f)$. 

In modulated time evolution, our time-dependent Hamiltonian becomes
\begin{align}
    \hat{H}_{\text{mod}}(t)=\tilde{\lambda}(t) \left[\hat{H}_A+B(t)\hat{H}_B\right].
    \label{eq:mod-ham}
\end{align}

As shown later, the functional forms of both fields are optimized in this work.  In our calculations, the time-evolution is approximated by a Trotter product formula~~\cite{trotter1959product,suzuki1991general}, yielding a time-evolved state
\begin{equation}
\begin{split}
\ket{\psi_f} = \prod_{j=1}^{N} e^{-i\hat{H}_{mod}(j\Delta t)} \ket{\psi_0}.
\label{eq:exp_time_evolution}
\end{split}
\end{equation}
Since we use a variational principle to determine \( \Delta t\, \tilde{\lambda}(t) := \lambda(t) \), we simply absorb the time step \( \Delta t \) into the definition of the modulation. The concrete optimization details, such as the choice of optimizer, are provided in Appendix~\ref{sec:optim numerics}.

\section{Modulated time evolution}

\subsection{Algorithm}
\begin{figure*}[htp]
    \begin{centering}
        \includegraphics[width=2\columnwidth]{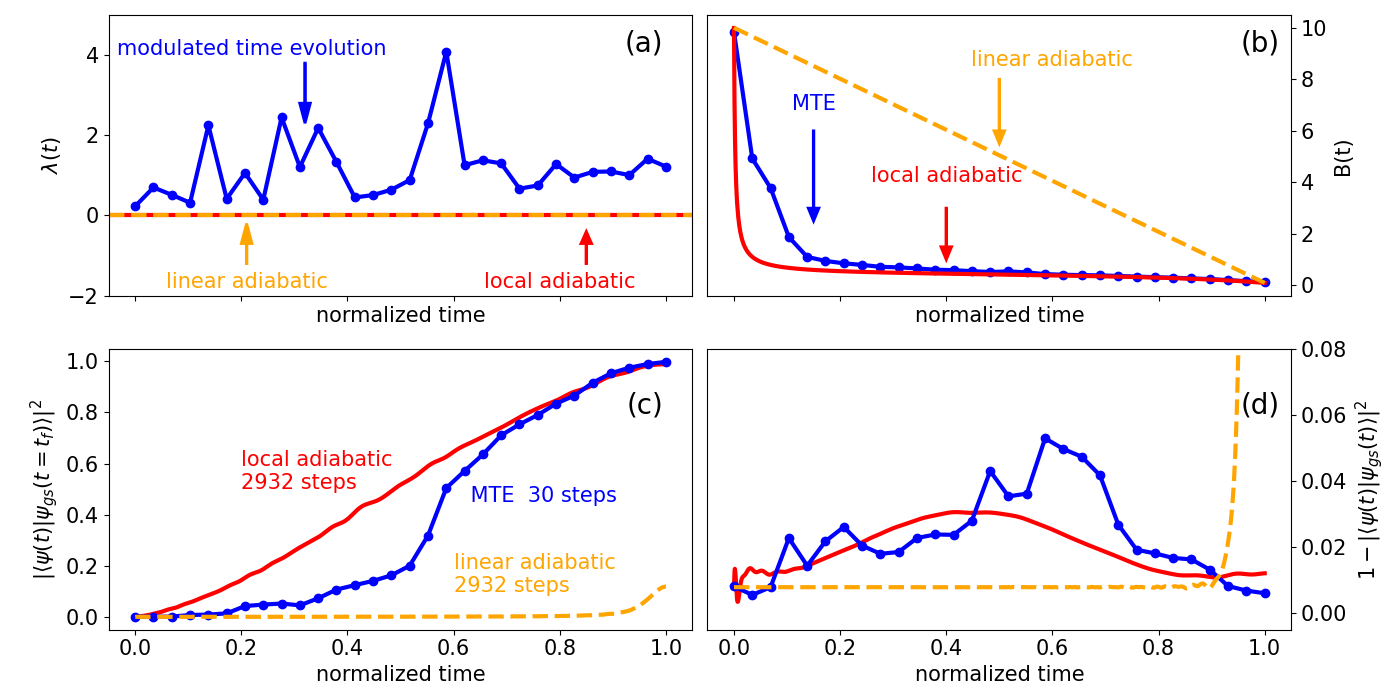}
    \end{centering}
    \caption{ Differences between the modulated time evolution (represented by solid blue line with dots) and the adiabatic time evolution (local one represented by red solid line and linear one represented by orange dashed line)  in the 12-site model, plotted in normalized time  $t/t_f$ . In modulated time evolution,  $t_f$  is not uniquely defined, so each step is plotted with uniform spacing, calculated as the reciprocal of the total number of steps. In the local adiabatic evolution, the total time is 29.32 with the adiabaticity parameter $  \rho = 10$, resulting in 2932 steps with  $\Delta t = 0.01$  (see Appendix~\ref{sec:LA field} for more details on local adiabatic ramp). The same number of steps is then chosen for linear adiabatic evolution. (a) The optimized scaling parameter $\lambda(t)$. In the adiabatic evolution cases, the scaling factor $\tilde{\lambda}(t)$ is set to one with $\Delta t = 0.01$, resulting in $\lambda(t) = 0.01$.  (b) The optimized field $B(t)$. (c) Target ground state fidelity. Note that linear adiabatic evolution performs poorly as it does not slow down when the energy gap is small, where most diabatic excitation occurs. (d) The instantaneous infidelity.
    \label{fig:lamda B plot}}
\end{figure*}

The open boundary condition long-range transverse-field Ising model is given by
\begin{equation}
\begin{split}
\hat{H}_0(t) = \sum_{i<j} J_{i,j} \hat{\sigma}_z^{(i)} \hat{\sigma}_z^{(j)} + B(t) \sum_{i} \hat{\sigma}_x^{(i)},
\label{eq:ising model}
\end{split}
\end{equation}
We denote the Hamiltonians as $\hat{H}_A = \sum_{i<j} J_{ij} \hat{\sigma}_z^{(i)} \hat{\sigma}_z^{(j)}$ and $\hat{H}_B = \sum_{i} \hat{\sigma}_x^{(i)}$, where $J_{ij}$ represents the Ising coupling strength between spins at lattice sites $i$ and $j$. In this study, we consider cases such as long-range antiferromagnetic couplings~\cite{koffel2012entanglement}, defined by $J_{ij} = 1/|i - j|$ for $i \neq j$, as well as spin-glass couplings, where the $J_{ij}$ values are independently drawn from a Gaussian distribution: $J_{ij} \sim \mathcal{N}(\mu = 0, \sigma^2 = 1)$, with mean $\mu$ and variance $\sigma^2$. We set the reduced Planck constant $\hbar = 1$. 

We also do not apply Kac scaling~\cite{kac1963van}, so that the coupling coefficients $J_{ij}$ match those used in the ion-trap experimental study of Ref.~\cite{pagano2020quantum}, enabling a direct discussion, which we elaborate on in Sec.~\ref{sec:QAOA}. To properly address the dependence of the energy gap as the system size grows—particularly since the omission of Kac scaling~\cite{kac1963van} can influence the extensive property—we include a discussion of energy gap dependence in the next section.

The time-dependent transverse magnetic field is denoted by $B(t)$, and $\hat{\sigma}_{\beta}^{(i)}$ is the Pauli spin operator acting on spin $i$ along the $\beta$ direction. The target ground state preparation of Eq.~(\ref{eq:ising model}) corresponds to a final ramp value of \( B(t_f) = 0.1 \). Note that, unlike the case with only nearest-neighbor interactions, which can be solved analytically via the Bogoliubov transformation~\cite{calabese2011quantum}, the ground state of the long-range model lacks a known analytical solution.

At first glance, both quantum annealing~\cite{kadowaki1998quantum,morita2008mathematical,brady2021optimal,hegde2023deep,susa2021variational,crosson2021prospects} and modulated time evolution aim to accelerate the time evolution process by allowing diabatic excitations, but modulated time evolution differs from typical quantum annealing in its control structure and time constraint. 
First, typical quantum annealing employs a control Hamiltonian of the form
\[
\hat{H}(t) = u(t)\hat{H}_A + (1 - u(t))\hat{H}_B,
\]
with boundary conditions \( u(0) = 0 \) and \( u(t_f) = 1 \) imposed to ensure practical implementation. This framework typically relies on a single control function \( u(t) \), whose optimal shape can, in principle, be determined using optimal control theory~\cite{brady2021optimal}.

Next, several prior works~\cite{hegde2022genetic,matsuura2021variationally}, have also explored the use of two separate control functions in quantum annealing. A key distinction in our approach is that we express \( \lambda(t) \) as a prefactor of the full Hamiltonian, rather than assigning separate control functions to \( \hat{H}_A \) and \( \hat{H}_B \) individually. This seemingly minor but key structural difference enables the construction of analytically motivated initial guesses for both the parameters and the functional forms of the control functions \( \lambda(t) \) and \( B(t) \). This formulation not only offers physical interpretation but also improves numerical performance, as it allows us to begin the optimization with well-informed initial guesses based on the physical roles of these fields. 

Finally, most time evolution strategies—including quantum annealing and fast-forward driving~\cite{brady2021optimal,petiziol2018fast,BukovReinforcement2018,gu2021fast}—impose a fixed total evolution time \( t_f \). In contrast, modulated time evolution does not uniquely define \( t_f \), as only the product \( \Delta t\, \tilde{\lambda}(t) \) is specified, without independent knowledge of \( \Delta t \).
Hence, in this work, without imposing total time \( t_f \) as a constraint, we focus on another equally important quantity: the number of time steps, \( N \). Given \( N \), the initial guesses for the two fields are specified as follows. For \( B(t) \), since the energy gap tends to be large when \( B \) is large and small when \( B \) is small, we adopt an exponentially decaying function as the initial guess:
\[
B(t_j) = B_0 \exp\left( -\frac{j \tau}{N-1} \right),
\]
where \( \tau = \log(B_0 / B_f) \), \( B_0 = 1 \), \( B_f = 0.1 \), and \( j \in [0, N-1] \) is a non-negative integer denoting the time step index. 

For $\lambda(t)$, we use perturbation theory to help us determine what a reasonable initial guess is. We start our system in the ground state of the initial Hamiltonian $\hat{H}_B$, and then express the subsequent time evolution in the form
\begin{equation}
 |\Psi(t) \rangle =  \sum_n c_n(t) |n(t)\rangle e^{i \theta_n (t) },
\end{equation}
with the dynamical phase $ \theta_n(t) =  -\int_0^t \tilde{\lambda}(\bar{t})E_n (\bar{t}) d\bar{t} $ and the instantaneous energy given by $\tilde{\lambda}(t)E_n(t)$; here, we have $\hat{H}_{\text{mod}}(t)|n(t)\rangle= \tilde{\lambda}(t)E_n(t)|n(t)\rangle$.

The instantaneous state amplitude satisfies the differential equation
\begin{align}
&\frac{d}{dt}c_m(t) =  -c_m(t)\langle m(t)|\dot{m}(t)\rangle \nonumber\\
& -\sum_{n\ne m}c_n(t)\frac{ \tilde{\lambda}(t) \dot{B}(t)\langle m(t)|\hat{H}_{B}|n(t)\rangle}{\tilde{\lambda}(t)\big (E_n(t)-E_m(t)\big )} 
e^{-i\big(\theta_n(t)-\theta_m(t)\big)}
\label{eq:inst_gs_formula}
\end{align}
The first term on the right-hand side is independent of $\tilde{\lambda}(t)$ because it depends solely on the eigenvectors and their evolution. Unlike $B(t)$, which can influence the eigenvectors through the Hamiltonian structure, $\tilde{\lambda}(t)$ cannot. We observe that the $\tilde{\lambda}(t)$ factors cancel out in both the numerator and denominator of the second term, leaving its dependence only through the dynamical phase $\theta_n(t)$.

Since $\tilde{\lambda}(t)$ enters as a modulation of the integral over energy, its influence is most significant when it is a rapidly oscillating field. Motivated by this observation, we initialize ${\lambda}(t)$ as a random array with values ranging from 1 to 2, aiming to qualitatively capture an oscillatory character. If prior knowledge of the energy spectrum were available, a more refined initialization could be used. However, in typical cases where such information is unavailable, the chosen range of 1 to 2 provides a practical balance—introducing sufficient modulation while keeping the magnitude moderate for potential experimental implementation.

\begin{figure}[htp]
\begin{centering}
    \includegraphics[width=0.9\columnwidth]{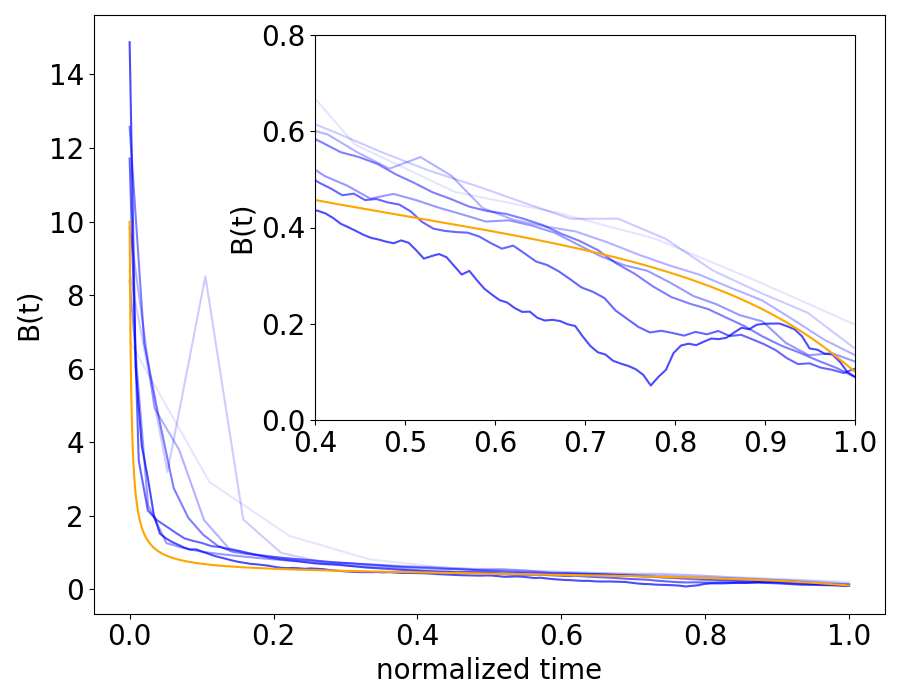}
\end{centering}
\caption{Optimal \( B(t) \) for a 12-site lattice is shown for varying numbers of steps \( N = 10, 20, 30, 40, 50, 80, 120 \). Darker lines correspond to higher layer counts. The inset shows magnifies the results for the optimal \( B(t) \) at later times. The orange solid line is the local adiabatic ramp.
\label{fig:la vs B}}
\end{figure}

In Fig.~\ref{fig:lamda B plot}, panels (a) and (b), we show the 30-step optimized $\lambda(t)$ and $B(t)$ for the 12-site model. While oscillations are clearly present in the optimized $\lambda(t)$, we observe no consistent pattern in their frequency or amplitude. This suggests that these oscillations can be primarily shaped by the choice of $N$ and the initial parameter guess, rather than being determined by a physical energy scale in the system. In contrast, the optimized $B(t)$ exhibits a smooth, convergent profile that closely resembles the local adiabatic ramp, as illustrated in Fig.~\ref{fig:la vs B} when plotted against normalized time for different values of $N$.

In Fig.~\ref{fig:lamda B plot}  panel (c) and (d), we present the fidelity as well as instantaneous infidelity as a function of normalized time, comparing it to the adiabatic evolution. In particular panel (d) highlights how the modulated time evolution enables the evolved state to deviate from the instantaneous ground state while ultimately eliminating diabatic excitations to achieve a high target ground state fidelity. Additionally, in the case of local adiabatic evolution shown in panel (d), there is also a return in the instantaneous ground state fidelity. However, this is attributed to finite-time evolution effects, as true adiabatic evolution—under infinitely slow processes—would result in a flat line at 1. More details are provided in the Appendix~\ref{sec:adiabatic _insta_info}, which demonstrates that the return amplitude of the instantaneous ground state fidelity can be significantly reduced by adopting a more adiabatic-like evolution setting. Interestingly, similar returning behavior was also observed in Ref.~\cite{kovalsky2023self}, which studied the dependence of Trotter error on total evolution time and time-step size.

\begin{figure}
\begin{centering}
\includegraphics[width=0.99\columnwidth]{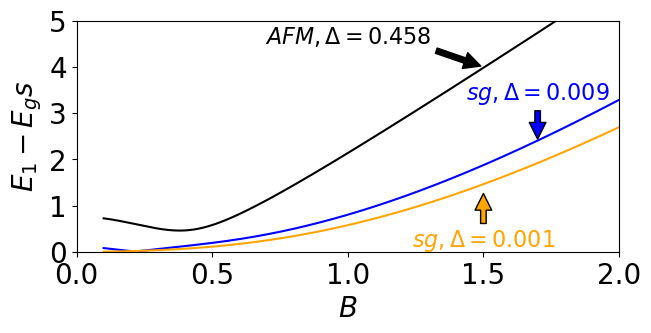}
    \end{centering}
    \caption{Energy gap between the first coupled excited state and the ground state for the antiferromagnetic (AFM) system and two different spin-glass (sg) systems, calculated on an 8-site lattice. Here, \( \Delta \) denotes the minimum energy gap value.
 \label{fig:energy gap afm sq}}
\end{figure}

In conclusion, the functional forms of the two control fields have become clear: \( \lambda(t) \) serves as an oscillating field, while \( B(t) \) resembles a local adiabatic ramp. Naturally, the evolution driven by \( B(t) \), or equivalently the transition between \( \hat{H}_B \) and \( \hat{H}_A \), follows a trajectory similar to local adiabatic evolution—transitioning more rapidly when the energy gap is large. However, due to the finite number of steps available in practice, diabatic excitations are unavoidable. This motivates the use of an oscillating field \( \lambda(t) \), which permits controlled diabatic excitations during the evolution, as long as they are largely suppressed by the end. For instance, in Fig.~\ref{fig:lamda B plot}(d), the instantaneous infidelity remains below 6\% in this case and below 20\% across all studied scenarios. Higher infidelity typically occurs with fewer time steps. Crucially, these diabatic excitations are mostly eliminated by the conclusion of the evolution.

In practical implementations, the potentially rapid oscillations in \( \lambda(t) \) may be challenging to realize on analog quantum simulators, such as ion-trap hardware. Nonetheless, as a basic study, the aim of this work is to identify an optimized theoretical control strategy. To bridge the gap between theory and experiment, we also examine the most constrained scenario in which \( \lambda(t) \) is held constant, as shown in Appendix~\ref{sec:constant mte}. Even under this simplification, the performance remains robust—achieving a fidelity of 0.99 with substantially fewer steps than adiabatic time evolution—when \( B(t) \) is chosen as an exponential decay, without any optimization.

\subsection{Energy gap dependence}

\begin{table}
    \centering
    \begin{tabular}{|c|c|c|c|}
        \hline
        \multirow{5}{*}[0.1em]{steps} & \multicolumn{1}{c|}{AFM} & \multicolumn{2}{c|}{spin glass} \\ \cline{2-4}
        & $ B_c = 0.38 $ & $ B_c = 0.215 $ & $ B_c = 0.1 $ \\ 
        & $ \Delta = 0.458 $ & $ \Delta = 0.009 $ & $ \Delta = 0.001 $ \\ \hline
        10 & $3.55 \times 10^{-3}$ & $5.05 \times 10^{-2}$ & $5.33 \times 10^{-2}$ \\ \hline
        20 & $2.67 \times 10^{-4}$ & $3.08 \times 10^{-2}$&$1.79 \times 10^{-2}$ \\ \hline
        50 & $3.04 \times 10^{-5}$ & $1.58 \times 10^{-2}$&$1.19 \times 10^{-2}$ \\ \hline
        80 & $5.57 \times 10^{-10}$ & $7.15 \times 10^{-3}$&$2.69 \times 10^{-3}$ \\ \hline
        200 & $\backslash$  &$9.22 \times 10^{-10}$ &$1.67 \times 10^{-8}$ \\ \hline
        300 &$\backslash$ &$\backslash$ &$3.35 \times 10^{-10}$ \\ \hline
    \end{tabular}
    \caption{1 - \( E/E_{\mathrm{gs}} \) in the 8-site model, where \( E \) is the final evolved energy. AFM refers to the antiferromagnetic couplings studied in this work, along with two different spin glass cases.
  \label{table:enegy_gap_vs_E} }
\end{table}


\begin{figure*}[htp]
\begin{centering}
\includegraphics[width=\textwidth]{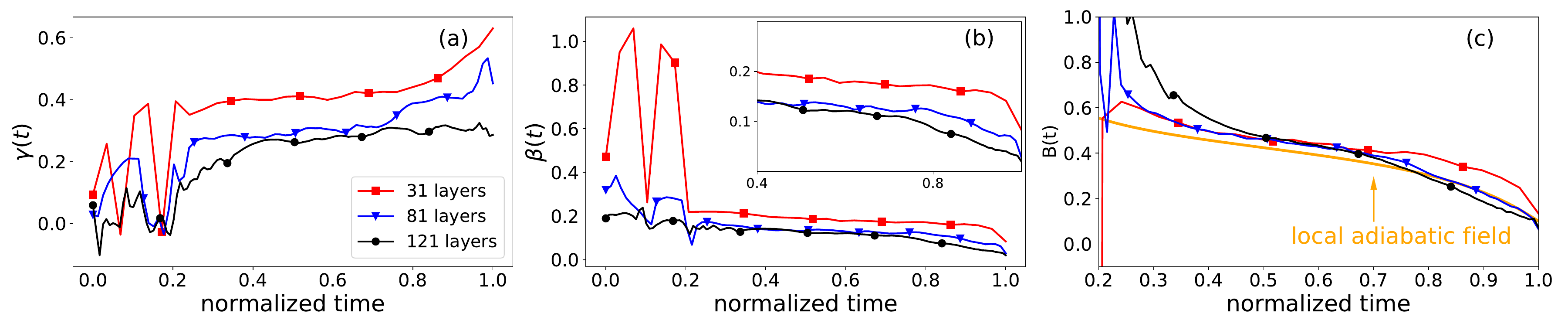}
    \end{centering}
    \caption{Optimized 12-site QAOA angles and its ratio versus normalized time. (a) QAOA angle $\gamma (t)$.  (b) QAOA angle  $\beta (t)$.  (c) The ratio  $\beta (t)$/$\gamma (t)$. The ratio compares with the local adiabatic field, which is plotted in the orange solid line.  Note, we count the last single term $\exp{-i \gamma\hat{H}_A/(2m)}$ as one layer, resulting in the total number of layers to be 31, 81 and 121.  
 \label{fig:QAOA-ratio}}
\end{figure*}
\begin{figure}
\begin{centering}
    \includegraphics[width=0.8\columnwidth]{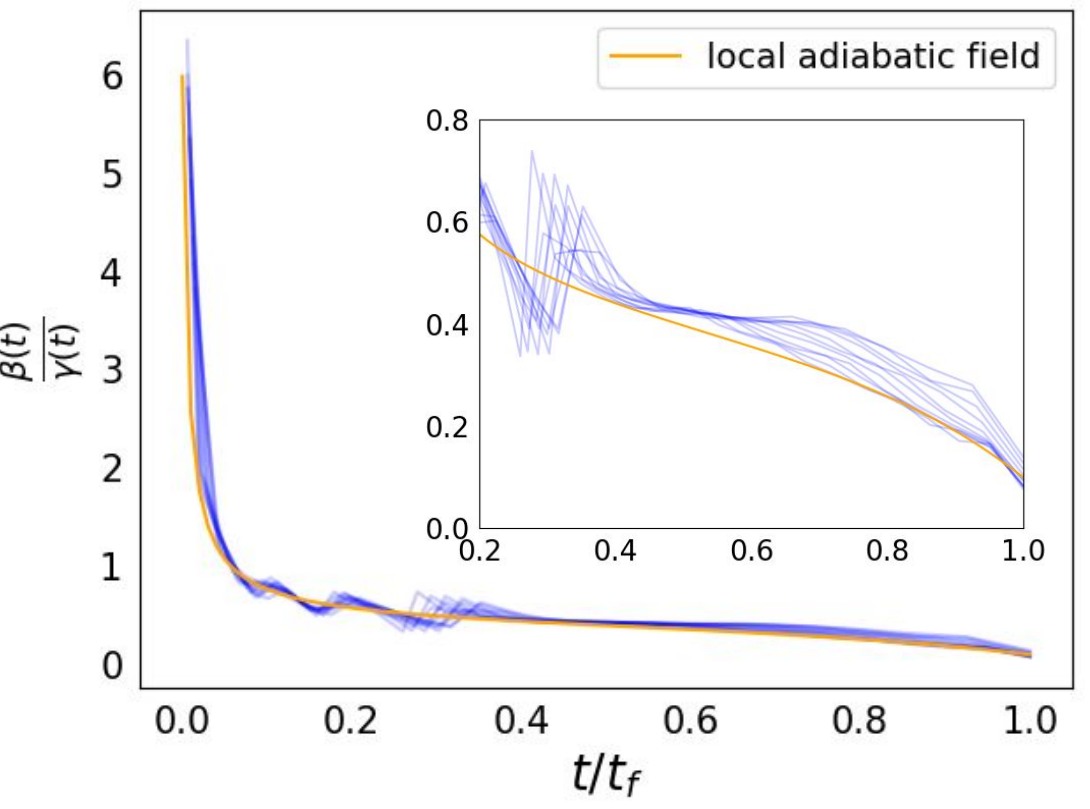}
\end{centering}
\caption{We compute the QAOA angle ratio using data from Ref.~\cite{pagano2020quantum} for the 8-site case. The plot displays the number of layers ranging from 20 to 30, with curves corresponding to higher \(p\) values shaded darker. The data is plotted against normalized time.
\label{fig:QAOA ratio umd}}
\end{figure}

In adiabatic evolution, it is known that the total evolution time scales as $\propto 1/\Delta^2$, where $\Delta$ is the minimum energy difference between the ground state and the first coupled excited state during the evolution~\cite{roland2002quantum}. As the system size increases, systems that become gapless at the critical point in the thermodynamic limit experience a continual decrease in the energy gap, posing significant challenges for adiabatic time evolution in large systems with narrow energy gaps. Hence, it is of particular interest to explore how modulated time evolution performs as the system size increases and the energy gap narrows.

Unlike the quantum Monte Carlo method~\cite{kaicher2023mean}, which can simulate systems with hundreds of spins, the classical simulation currently employed relies on the matrix exponential of the time evolution operator. The time complexity of the simulation scales as high as $O(n^3)$~\cite{Higham2005the} for a general dense matrix, where $n$ is the dimension of the matrix in Eq.~(\ref{eq:exp_time_evolution}). As a result, the simulation is restricted to small lattice sizes, preventing scaling analysis for larger systems. Instead, we focus on spin-glass systems with substantially smaller energy gaps to emphasize the weak energy gap dependence of this method.

Using the 8-site model as an example, we generate two different energy gap scenarios for the spin-glass model, as shown in Fig.~\ref{fig:energy gap afm sq}. In Table~\ref{table:enegy_gap_vs_E}, we find that even when the minimum energy gap value decreases by more than two orders of magnitude, modulated time evolution requires only a three- to four-fold increase in the number of time steps to maintain comparable relative energy error. In contrast, adiabatic scaling would predict an increase in time steps by approximately four orders of magnitude. This result highlights the weak dependence of modulated time evolution on the energy gap and underscores the speed up due to the benefits of allowing diabatic excitations when the gap is exceedingly small, where such excitations are nearly unavoidable.



\section{Connection to QAOA }
\label{sec:QAOA}

The quantum approximate optimization algorithm~\cite{farhi2014quantum,farhi2016quantum} is a simple to implement yet powerful  protocol~~\cite{pagano2020quantum,shaydulin2023evidence,Lotshaw2023appromixate}, which involves $p$ layers, with each layer corresponding to the application of a mixing Hamiltonian $\hat{H}_{B}$ (weighted by the angle $\beta_j$) and a problem Hamiltonian $\hat{H}_{A}$ (weighted by the angle $\gamma_j$), with $j$ denoting the layer (from 1 to $p$).  The QAOA then has the following variational ansatz for the optimization problem:
\begin{align}
    |\psi_f\rangle&=e^{-i\beta_p\hat{H}_{B}}e^{-i\gamma_p\hat{H}_{A}}e^{-i\beta_{p-1}\hat{H}_{\text{B}}}e^{-i\gamma_{p-1}\hat{H}_{\text{A}}}\cdots \nonumber\\
    &\times e^{-i\beta_1\hat{H}_{\text{B}}}e^{-i\gamma_1\hat{H}_{\text{A}}}|\psi_0\rangle,
    \label{eq:qaoa}
\end{align}
which constructs the final ansatz state $|\psi_f\rangle$ in terms of the initial state $|\psi_0\rangle$. The initial state $| \psi_0\rangle $ is also the ground state of $\hat{H}_B$. 

Many studies have been performed to understand and improve QAOA  ~\cite{kremenetski2023quantum, rosa2021Universal,lotshaw2021empirical,streif2019Comparison,sack2021quantum,diezvalle2023quantum,vikst2020Applying,willsch2022gpu,lubinski2023optimization,yao2022monte}, link QAOA to optimal control theory~~\cite{kocia2021behavior,brady2021optimal,an2022quantum}, or the counter-diabatic method ~~\cite{wurtz2022counterdiabaticity,Wurtz2021Maxcut,chandarana2022digitized} in an effort to reveal the underlying nature of QAOA angles, with the aim of finding a more efficient strategy to optimize the angles. 

To first obtain the QAOA angles, we aim to use the parameters obtained from modulated time evolution as an initial guess for QAOA, with the goal of achieving the best possible QAOA performance. We begin by rewriting the modulated time evolution expression in Eq.~(\ref{eq:exp_time_evolution}) into the QAOA form, as given in Eq.~(\ref{eq:qaoa}). This is achieved by applying a second-order Trotter product formula, which transforms a single time-evolution layer from our earlier calculations into \( m \) layers in the QAOA expression, as follows:

\begin{align}
e^{-i(\lambda \hat{H}_A + \lambda B \hat{H}_B)} &=
e^{-i \frac{\lambda}{2m}\hat{H}_A} \left(e^{-i \frac{\lambda}{m} \hat{H}_A} e^{-i \frac{\lambda B}{m} \hat{H}_B}\right)^m \nonumber \\
&\times e^{-i \frac{\lambda}{2m}\hat{H}_A} + O\left(\frac{1}{m^2}\right).
\label{eq:trotter_expand}
\end{align}
This Trotter formula translates the modulated time evolution form into the QAOA form. For each time step in the modulated time evolution, we select a separate number of Trotter terms, $m$, such that the final energy from the QAOA form with initial parameters remains within 30\% of the energy obtained from the modulated time evolution form. The 30\% margin is chosen empirically: if the margin is too narrow, it results in an excessive number of layers in the QAOA form. Conversely, if the margin is too wide, the energy from the QAOA form deviates significantly from the pre-Trotter result, making the resulting angles unhelpful, as arbitrary angles could potentially produce similar energy values. This process typically maps the number of layers from $N$ in modulated time evolution to approximately $4N$ in QAOA. More details are provided in the Appendix~\ref{sec:error ratio}.

\begin{figure}[htp]
    \begin{centering}
        \includegraphics[width=\columnwidth]{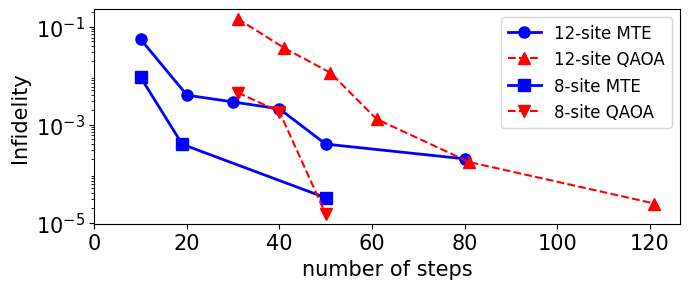}
    \end{centering}
    \caption{Target ground state infidelity verus number of steps from using modulated time evolution and QAOA in the 8-site and 12-site models.
    \label{fig:mod vs QAOA performance}}
\end{figure}

With this informed initialization, we typically observe significantly improved QAOA angles compared to those obtained from an uninformed guess. In Ref.~\cite{pagano2020quantum}, QAOA angles were optimized for up to 30 layers, achieving a target ground state fidelity around 0.99. Here, under the same setup, and as shown in Fig.~\ref{fig:mod vs QAOA performance}, QAOA can achieve even higher accuracy---reaching infidelities on the order of $10^{-5}$---when more layers are allowed. 

We emphasize that this procedure is not intended as a practical method for generating QAOA initialization angles, since it requires a full optimization of the modulated time evolution beforehand. However, it enables us to obtain the best available QAOA angles, allowing for a study to understand the underlying structure.

In this work, our attempt to understand QAOA is guided by the perspective of modulated time evolution. To make this connection more transparent, we begin by rewriting a single \( i \)th layer of the QAOA expression into a single exponential form, which more closely resembles the modulated time evolution formulation, as shown below:

\begin{equation}
e^{i \gamma_i(t) \hat{H}_{A}} e^{i \beta_i(t) \hat{H}_{B}} \\
\approx e^{i \gamma_i(t)(\hat{H}_{A} + \frac{\beta_i(t)}{\gamma_i(t)} \hat{H}_{B} - \frac{i \beta_i(t)}{2} \left[\hat{H}_{A}, \hat{H}_{B}\right] + \dots)}.
\label{eq:bch_qaoa}
\end{equation}

Focusing on the dominant terms, $\gamma(t)$ and $\beta(t)/\gamma(t)$ (since most QAOA angles are on the order of $10^{-1}$, making higher-order terms less important), we interpret $\gamma(t)$ as a prefactor or an effective time step size—particularly when viewing the evolution as a discretized time process rather than as modulation of the Hamiltonian. Accordingly, we define the total evolution time as the sum of the QAOA angles \( \gamma_i \), i.e., \( t_{\text{total}} = \sum_{i=1}^{p} |\gamma_i| \), rather than summing over both \( \gamma_i \) and \( \beta_i \) as is commonly done~\cite{zhou2020quantum}. In this context, the angle ratio \( \beta(t)/\gamma(t) \) can be interpreted as a time-dependent ramp, analogous to \( B(t) \) in modulated time evolution. 

Using the time definition introduced above, Fig.~\ref{fig:QAOA-ratio} plots the QAOA angles for several layer depths in the 12-site model. A clear pattern emerges: once the early-time oscillations subside, the ratio \(\beta(t)/\gamma(t)\) coincides with the locally adiabatic ramp. Large-amplitude oscillations can appear during the first \(\sim 0.2\) of the normalized time when an individual angle \(\gamma_i\) is very close to zero (positive or negative), which inflates the ratio \(\beta_i/\gamma_i\). The occurrence of such near-zero \(\gamma_i\) values depends on the chosen layer depth and the initial parameter guess. Although a vanishing \(\gamma_i\) contributes negligibly to the evolution—and could be removed by imposing a small threshold—we retain these points to present a complete data set.

We attribute the early-time oscillations caused by the optimizer getting trapped in different local minima rather than to any specific physical mechanism; such oscillations depend on the choice of optimizer and initial conditions. Independent data from Ref.~\cite{pagano2020quantum} (Fig.~\ref{fig:QAOA ratio umd})—obtained with a different optimization protocol—exhibit the same long-time alignment with the locally adiabatic ramp. In that data set, no large early-time spikes are present; instead, only small-amplitude oscillations appear between normalized times 0.2 and 0.4. For the 8-site system with 30 layers, the fidelity difference between the two angle sets is minor, on the order of \(10^{-2}\), confirming the practical effectiveness of both.

With both QAOA data and modulated time evolution data in hand, Fig.~\ref{fig:mod vs QAOA performance} compares the ground state infidelity versus the number of time steps \( N \) for 8- and 12-site spin models. Across the full range explored, modulated time evolution consistently matches—and in some regimes even surpasses—the performance of QAOA. These results highlight modulated time evolution as an alternative approach for near-term analog quantum algorithms.

\section{Conclusion}

We have introduced modulated time evolution, an intuitive strategy that accelerates quantum dynamics by permitting controlled diabatic excitations, thereby reducing the number of time steps while still achieving high ground state fidelity. This framework uncovers a unifying structure shared with QAOA: in both protocols the evolution is steered by an approximately local-adiabatic ramp. In QAOA that guiding field is encoded in the angle ratio \( \beta(t)/\gamma(t) \); in modulated time evolution it appears explicitly as the transverse field \( B(t) \).

We hope this study offers a practical bridge among adiabatic, diabatic, and variational paradigms, providing the community with a another route toward fast, high-fidelity quantum state preparation.
\section{Acknowledgments}

We thank Aniruddha Bapat and Alexey Gorshkov for providing the QAOA data from Ref.~\cite{pagano2020quantum}. We acknowledge helpful discussions with Shuchen Zhu about Trotter product formula errors and Efekan Kökcü about time evolution in spin models. This work was supported by the Department of Energy, Office of Basic Energy Sciences, Division of Materials Sciences and Engineering under grant no. DE-SC0023231. J.K.F. was also supported by the McDevitt bequest at Georgetown.

\section{Data Availability}
The data that support the findings of this article as well as the python code that run the calculations are openly available at~\cite{data}.

\bibliography{bibliography.bib}

\appendix

\renewcommand{\theequation}{A\arabic{equation}}
\setcounter{equation}{0}

\renewcommand{\thefigure}{A\arabic{figure}}
\setcounter{figure}{0}
\renewcommand{\thefigure}{A\arabic{figure}}
\renewcommand{\theHfigure}{A\arabic{figure}}

\section{Numerics for the optimization process}
\label{sec:optim numerics}

In this study, the cost function is the final evolved energy, 
\[
E = \langle \psi_f | \hat{H}_0(t{=}t_f) | \psi_f \rangle,
\]
evaluated with respect to the parameters \( \lambda(t) \) and \( B(t) \). To best locate the local minimum, we adopt an optimization strategy in which all parameters---both \( \lambda(t) \) and \( B(t) \)---are optimized simultaneously, in contrast to a step-by-step approach. Although more computationally expensive, this strategy is widely used in accurate wavefunction ansatz-based quantum algorithms~\cite{grimsley2019adaptive,motta2023bridging,Cade2020Strategy}. In this work, our choice is motivated by the goal of identifying a short evolution trajectory that minimizes only the final energy. In this setting, it is not necessary to remain close to the instantaneous ground state throughout the evolution; only the final energy outcome is relevant. In future work, to improve practicality, a step-by-step scheme could be explored, and comparing its performance with full-trajectory optimization presents an important direction.

In this work, we use the BFGS optimizer from the SciPy package~\cite{2020SciPy-NMeth} on a CPU for the 8-site model. For the 12-site model, we switch to the L-BFGS optimizer from the PyTorch package~\cite{paszke2017automatic}, which supports GPU acceleration. The hyperparameter settings, such as the learning rate and convergence criterion—defined as the L2 norm of the gradient falling below \(10^{-5}\)—are kept the same as those used for the 8-site model. The key distinction is that L-BFGS approximates the inverse Hessian using information from only the most recent 100 iterations (as set in our implementation), rather than constructing it from the full optimization history as in the standard BFGS method.

\begin{figure}
\begin{centering}
\includegraphics[width=\columnwidth]{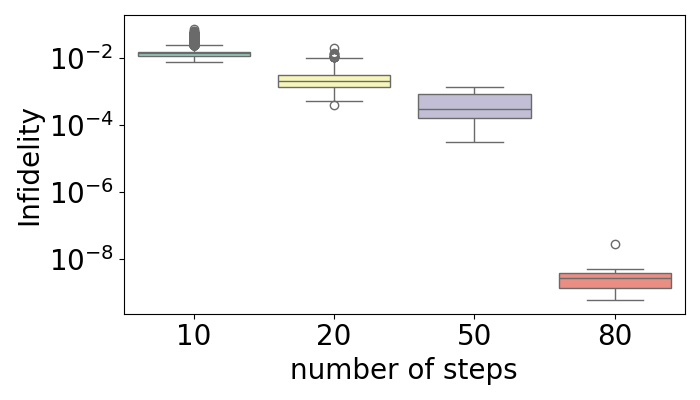}
\end{centering}
\caption{ Box plot of the fidelity versus the number of steps for the 8-site model. The number of runs for 10, 20, 50, and 80 layers are 6813 (6418) total (successful), 469 (360), 25 (21), and 22 (22), respectively. 
\label{fig:fidelity_boxplot}}
\end{figure}

In Fig.~\ref{fig:fidelity_boxplot}, we present a boxplot~\cite{velleman1981applications} of the 8-site optimization results, illustrating the convergence performance with various initial guesses over different numbers of steps.

In the boxplot, the line inside the box represents the median. The lower edge is the first quartile, the median of the lower half (25th percentile). The upper edge is the third quartile, the median of the upper half (75th percentile). Whiskers extend to the smallest and largest values within 1.5 times of the interquartile range. Points outside this range are outliers, shown as individual circles.

From this plot, it is evident that the fidelity is fairly consistent across different initial guesses, indicating that the optimization process does not heavily depend on the initial guess. Approximately 20\% (or less) of the runs fail to converge due to the inability of the optimizer to continue to successfully approximate the gradient. Even for failed cases, the final optimization results are usually promising, but we do not include them in the boxplots.

\begin{figure}
\begin{centering}
    \includegraphics[width=\columnwidth]{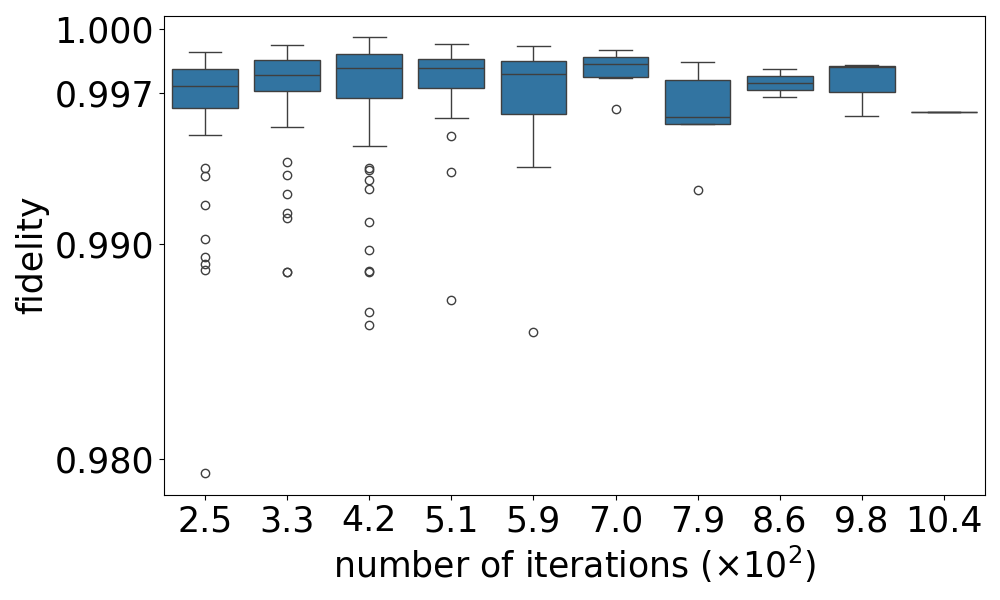}
\end{centering}
\caption{
Box plot for the 8-site model with 20 steps. We only include cases that successfully converged. The plot shows  the number of iterations versus converged fidelity.
\label{fig:f_vs_iternation}}
\end{figure}

In Fig.\ref{fig:f_vs_iternation}, we present a 8-site data boxplot showing the number of iterations versus fidelity in a 20-step optimization, using data from the 6418 convergence cases (out of 6813 tries). The number of iterations refers to how many times the optimizer updates $\lambda(t)$ and $B(t)$. The plot reveals that even though the optimizer sometimes requires significantly more iterations to converge, the median fidelity remains fairly consistent. This suggests the existence of many different local minima with similar energy levels but varying difficulties in finding them. Overall, it takes a few 100's of iterations to complete the optimization. In this case, it requires about 10 energy measurements per iteration.

\renewcommand{\theequation}{B\arabic{equation}}
\setcounter{equation}{0}

\renewcommand{\thefigure}{B\arabic{figure}}
\setcounter{figure}{0}
\renewcommand{\thefigure}{B\arabic{figure}}
\renewcommand{\theHfigure}{B\arabic{figure}}

\section{local adiabatic ramp construction and its comparison to the optimal B(t) }
\label{sec:LA field}
The local adiabatic ramp is chosen in such a way that the diabatic excitation out of the ground state is uniform for each time interval~\cite{roland2002quantum,richerme2013experimental}. The Landau-Zener problem tells us that the diabatic excitation depends on the instantaneous energy gap between the ground state and the first excited state of the same symmetry. We call this energy gap $\Delta(B(t))$, or more simply $\Delta(B)$.

\begin{figure}
\begin{centering}
    \includegraphics[width=0.9\columnwidth]{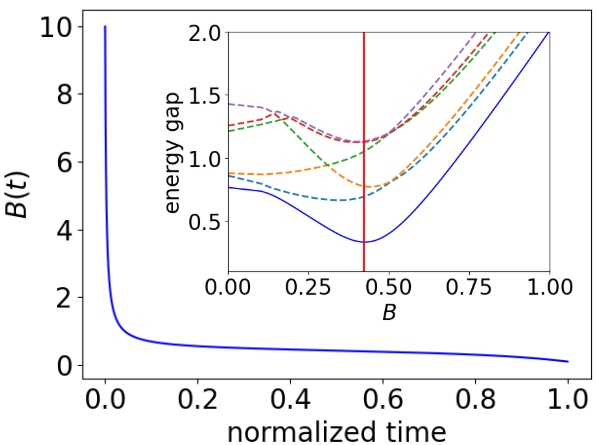}
\end{centering}
\caption{The local adiabatic ramp for an 12-site lattice starting from $B_{\text{max}}=10$ to $B_{\text{min}}=0.1$ (the same as in Fig.~1). Inset: the blue solid line represents $\Delta(B)$, illustrating the energy gap as a function of the magnetic field for the first coupled excited state. The dashed lines are for higher coupled excited states (not all excited states are shown here). The vertical red line marks the critical magnetic field with the minimum energy gap. 
\label{fig:gap}}
\end{figure}

Because the solution to the Landau-Zener problem tells us that the ratio of the square of the gap to the rate which we evolve the Hamiltonian determines the diabatic excitation, we define an adiabaticity parameter $\rho$, by
\begin{equation}
\begin{split}
\rho = \abs{ \frac{\Delta^2(B)}{\dot{B}(t)}     }
\end{split}
\end{equation}
and the local adiabatic ramp is determined by the function $B(t)$ that keeps $\rho$ constant over each time interval. This gives us
\begin{equation}
\begin{split}
t =\int_0^t\,d\bar{t} = \rho \int_{B_0}^{B} \frac{1}{\Delta^2(\bar{B})} \,d\bar{B},
\end{split}
\end{equation}
where we must pick the initial and current magnetic fields to determine the current time. By stepping the current magnetic field, we use the integral to generate the profile $t(B)$ (which is a monotonic increasing function) and then invert it to find the local adiabatic ramp $B(t)$ for the evolution to the final time $t_f$ (determined by the integral with an upper limit of $B_{end}$) and the adiabaticity parameter. An example for 12-site transverse field Ising model with $J_{ij} = 1/|i - j|^{-1}$ for $i\ne j$  is shown in Fig.~\ref{fig:gap}.

The instantaneous gap $\Delta(B)$ is calculated in the symmetry sector of the ground state by an exact diagonalization calculation. There are two parity symmetries for the transverse-field Ising model---a spin reflection parity and a spatial inversion parity. The spin-reflection parity is the eigenvalue of the ground state with respect to the spin-reflection operator, which takes $\sigma_x^{(i)}\to \sigma_x^{(i)}$, $\sigma_y^{(i)}\to -\sigma_y^{(i)}$, and $\sigma_z^{(i)}\to -\sigma_z^{(i)}$. The second symmetry is a spatial inversion symmetry. If we have $N$ lattice sites (with $i=1,~2,~\cdots,~N$), then the transformation $i\to N+1-i$ is a symmetry of the Hamiltonian because $J_{N+1-i,N+1-j}=J_{i,j}$. For  8(12)-sites and the antiferromagnetic case, the ground state is in the even-even sector. This reduces the Hamiltonian size from a dimension 256(4096) matrix to a dimension 72(1056) matrix and it allows us to easily identify the energy gap to the state with the same symmetry sector.

In Fig. \ref{fig:la vs B}, the optimal $B(t)$ with different numbers of steps is plotted along with the local adiabatic ramp on a normalized time axis. It is suggestive that the optimal $B(t)$ resembles the local adiabatic ramp more and more as the number of steps increases. 

\renewcommand{\theequation}{B\arabic{equation}}
\setcounter{equation}{0}

\renewcommand{\thefigure}{B\arabic{figure}}
\setcounter{figure}{0}
\renewcommand{\thefigure}{B\arabic{figure}}
\renewcommand{\theHfigure}{B\arabic{figure}}

\renewcommand{\theequation}{C\arabic{equation}}
\setcounter{equation}{0}

\renewcommand{\thefigure}{C\arabic{figure}}
\setcounter{figure}{0}
\renewcommand{\thefigure}{C\arabic{figure}}
\renewcommand{\theHfigure}{C\arabic{figure}}

\section{ Instantaneous fidelity versus B(t) in adiabatic time evolution}
\label{sec:adiabatic _insta_info}

We present additional details on the instantaneous ground state fidelity throughout the local adiabatic process, using the magnetic field $B(t)$ as the $x$-axis instead of the normalized time axis discussed in the main text.

\begin{figure}
\begin{centering}
    \includegraphics[width=\columnwidth]{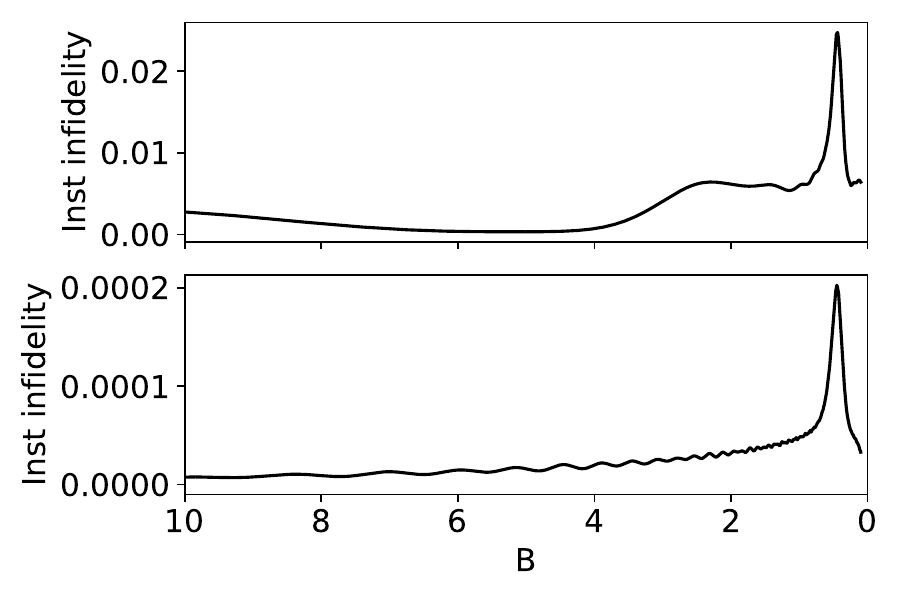}
\end{centering}
\caption{Instantaneous ground state infidelity during two local adiabatic ramp evolutions. The top panel corresponds to the configuration used in Fig.~\ref{fig:lamda B plot}, while the bottom panel represents a more adiabatic-like setting.
    \label{fig:main_text_la}}
\end{figure}

In Fig.~\ref{fig:main_text_la} top panel, the simulation parameters are defined as follows: $\rho = 10$, yielding a total simulation time $t_f = 29.32$. The initial magnetic field $B(t)$ is set to $B_0 = 10$, with a time step $dt = 0.01$. The system is initialized in the ground state of $\hat{H}_B$. However, since this state is not the true ground state at $B_0 = 10$, this setup introduces diabatic excitations from the start. In this configuration, we observe returns in the instantaneous fidelity, prompting an investigation into whether adiabatic evolution alone can account for this behavior. This is particularly intriguing because local adiabatic evolution typically introduces diabatic excitations at each step without mechanisms to eliminate them, yet such returns are observed here.

To further explore this, we present an additional case, shown in the bottom panel of Fig.~\ref{fig:main_text_la}, under modified parameters: $\rho = 100$, resulting in a total simulation time $t_f = 293.7$. The initial magnetic field $B(t)$ is set to $B_0 = 50$, with a smaller time step $dt = 0.001$. The system is initialized in the ground state of $\hat{H}_A + B_0 \cdot \hat{H}_B$, ensuring an initial state with virtually no diabatic excitations, as required by the adiabatic assumption. In this case, the same return pattern in the instantaneous ground state fidelity is observed, but with an amplitude smaller by a factor of 100.

\renewcommand{\theequation}{D\arabic{equation}}
\setcounter{equation}{0}

\renewcommand{\thefigure}{D\arabic{figure}}
\setcounter{figure}{0}
\renewcommand{\thefigure}{D\arabic{figure}}
\renewcommand{\theHfigure}{D\arabic{figure}}

\section{\texorpdfstring{Constant $\lambda_0$ time evolution}{Constant lambda0 time evolution}}
\label{sec:constant mte}
In the main text, there is no bound on $\lambda(t)$, but for practical reasons, one might impose a bounded range for $\lambda(t)$ or apply other specific settings. To demonstrate that this does not significantly affect modulated time evolution—aside from potentially requiring more steps—we present the most extreme case: reducing $\lambda(t)$ to a single parameter, $\lambda(t) = \lambda_0$.

\begin{figure}[htp]
\begin{centering}
    \includegraphics[width=\columnwidth]{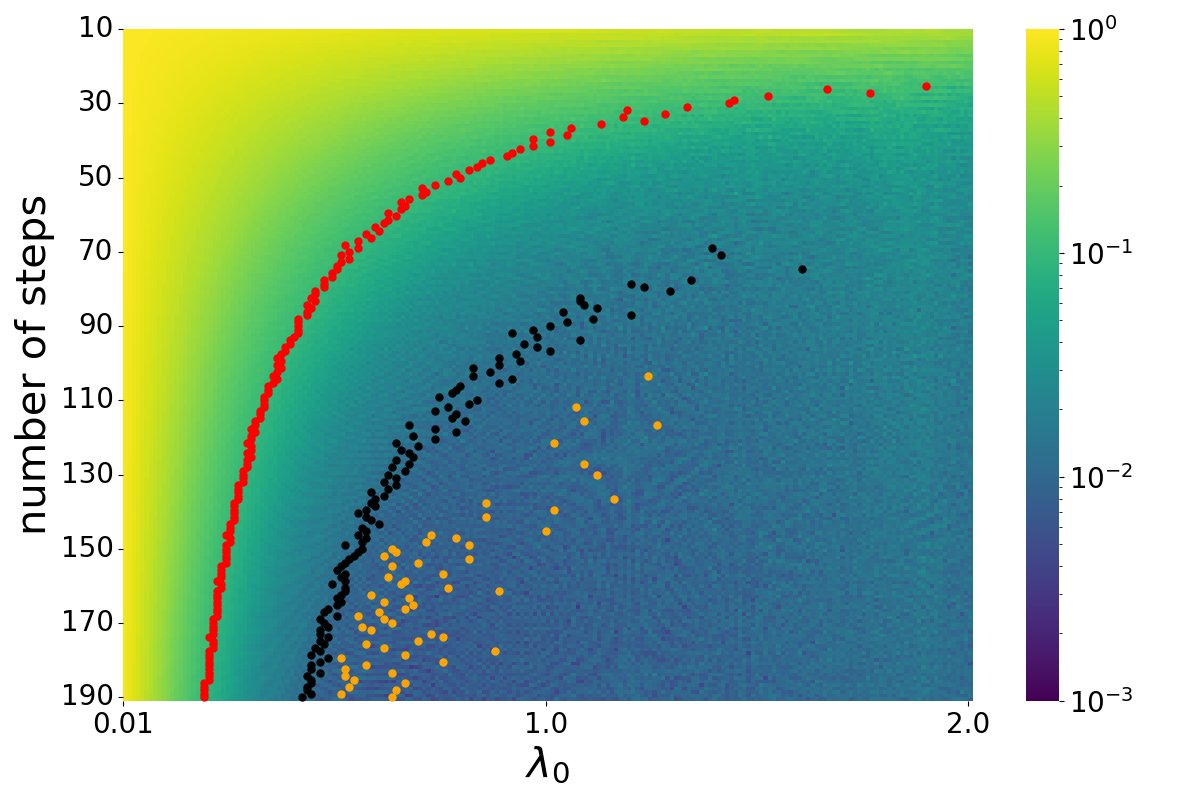}
\end{centering}
\caption{Infidelity heatmap of a constant $\lambda_0$-modulated time evolution for an 8-site model. For each step count, the dots highlights the minimum $\lambda_0$ values required to reach specific infidelity thresholds: 0.1 (red), 0.01 (black), and 0.005 (orange). 
\label{fig:constant lambda}}
\end{figure}

In Fig.~\ref{fig:constant lambda}, we present results using a single constant parameter $\lambda_0$. The data are displayed as a heatmap showing the target ground state infidelity across varying values of $\lambda_0$ and the number of steps, with no optimization of $B(t)$ needed; $B(t)$ is set as an exponentially decreasing field with $B_0 = 10$ and $B_f = 0.1$. It is shown that a wide range of $\lambda_0$ and step-count combinations yield fidelities of at least 0.99, although the highest fidelity we achieved was only 0.997. By contrast, modulated time evolution achieves 0.99 fidelity with an order of magnitude fewer steps.

We note that using a constant $\lambda$ also introduces a return mechanism, and in this case, we observe at most a 5\% deviation (infidelity) from the instantaneous ground state.

\renewcommand{\theequation}{E\arabic{equation}}
\setcounter{equation}{0}

\renewcommand{\thefigure}{E\arabic{figure}}
\setcounter{figure}{0}
\renewcommand{\thefigure}{E\arabic{figure}}
\renewcommand{\theHfigure}{E\arabic{figure}}

\section{Trotterization of modulated time evolution}
\label{sec:error ratio}
In Eq.~(\ref{eq:trotter_expand}), we show the  Trotter formula that translates the modulated time evolution form to a QAOA form. This requires us to approximate a single step in the modulated time evolution by $m$ Trotter steps for the QAOA form. 

We begin with the Hadamard lemma,
\begin{equation}
e^{\hat{A}}\hat{B}e^{-\hat{A}} \approx \hat{B} + [\hat{A},\hat{B}] + \frac{1}{2}[\hat{A}, [\hat{A},\hat{B}]]+\cdots,
\end{equation}
and apply it to the expression
\begin{equation}
e^{i \frac{\gamma}{2 m}\hat{H}_{A}} \left(e^{-i \frac{\gamma}{m} \hat{H}_{A}} e^{-i \frac{\beta}{m} \hat{H}_{B}}\right)^m  e^{-i \frac{\gamma}{2 m}\hat{H}_{A} },
\end{equation}
with $\hat{A}=i\tfrac{\gamma}{2m}\hat{H}_{A}$ and $\hat{B}$ the operator inside the parenthesis. We use the QAOA notation of  $\gamma$ and $\beta$ multiplying the problem and mixer Hamiltonians. 

Because of the similarity transformation, the outermost exponential factor goes into the powers inside the parenthesis and then into the argument of the function inside the parenthesis, so we need to evaluate
\begin{equation}
\left(e^{-i \frac{\gamma}{m} \hat{H}_{A}} \exp\left[ e^{i \frac{\gamma}{2 m}\hat{H}_{A}}\left (-i \frac{\beta}{m}\hat{H}_{B}\right ) e^{-i \frac{\gamma}{2 m}\hat{H}_{A}}\right]\right)^m,
\end{equation}
with the Hadamard lemma. Note that this acts only on the operator $\hat{H}_{B}$ because $\hat{H}_{A}$ commutes with itself. Expanding the Hadamard lemma out to second order gives us
\begin{align}
\Big (e^{-i \frac{\gamma}{m} \hat{H}_{A}} &\exp\Big \{-i \frac{\beta}{m}  \big (\hat{H}_{B}+i\frac{\gamma}{2m}[\hat{H}_{A},\hat{H}_{B}]\nonumber\\
&-\frac{\gamma^2}{8m^2}[\hat{H}_{A},[\hat{H}_{A},\hat{H}_{B}]]+\cdots\big)\Big\}\Big )^m.
\end{align}
Next, we use the Baker-Campbell-Hausdorff formula
\begin{equation}
e^{\hat{A}}e^{\hat{B}}=e^{\hat{A}+\hat{B}+\frac{1}{2}[\hat{A},\hat{B}]+\frac{1}{12}[\hat{A},[\hat{A},\hat{B}]]+\frac{1}{12}[\hat{B},[\hat{B},\hat{A}]]+\cdots},
\end{equation}
for each factor inside the parenthesis to put all terms into one exponent. Here we have 
\begin{equation}
    \hat{A}=-\frac{i\gamma}{m}\hat{H}_{A},
\end{equation} 
and 
\begin{align}
    \hat{B}=&-i\frac{\beta}{m}\Big (\hat{H}_{B}+i\frac{\gamma}{2m}[\hat{H}_{A},\hat{H}_{B}]\nonumber\\
    &-\frac{\gamma^2}{8m^2}[\hat{H}_{A},[\hat{H}_{A},\hat{H}_{B}]]\Big ),
\end{align} 
and we keep only the terms up to two-fold nested commutators. This gives
\begin{align}
    &\Big(\exp\Big\{-\frac{i}{m}(\gamma \hat{H}_{A}+{\beta}\hat{H}_{B})\nonumber\\
    &~~~~~~~~~~-\frac{i\beta \gamma^2}{24 m^3}[\hat{H}_{A},[\hat{H}_{A}, \hat{H}_{B}]]\nonumber\\
    &~~~~~~~~~~+\frac{i\beta^2\gamma }{12 m^3}[\hat{H}_{B},[\hat{H}_{B}, \hat{H}_{A}]]\Big\}\Big)^m.
\end{align}
Note how the first order commutator terms cancel out, leaving just the second order nested commutator terms, which come in with an additional power of $m$ in the denominator. This is the advantage of using this form.
Because all of the terms in the exponent are now the same, taking the $m^{th}$ power just requires us to multiply the exponent by $m$, to give
\begin{align}
    &\exp\Big \{-i\gamma \hat{H}_{A}-i{\beta}\hat{H}_{B}\nonumber\\
    &~~~~~~~~-\frac{i \beta\gamma^2}{24 m^2}[\hat{H}_{A},[\hat{H}_{A}, \hat{H}_{B}]]\nonumber\\
    &~~~~~~~~+\frac{i \beta^2\gamma}{12 m^2}[\hat{H}_{B},[\hat{H}_{B}, \hat{H}_{A}]]\Big\}. 
\end{align}
Hence, the error of this form is of order $1/m^2$.

We define an error ratio $\zeta$ to be
\begin{equation}
\zeta =\max \left (\frac{\abs{\frac{\gamma^2 \beta}{24 m^2}}}{\abs{\gamma} +\abs{\beta}},\frac{\abs{\frac{\gamma \beta^2}{12 m^2}}}{\abs{\gamma} +\abs{\beta}}\right ).
\end{equation}

This parameter determines the value of \( m \) needed to ensure that the energy difference before and after Trotterization is within 30\%. Unlike other works that provide full Trotter error analysis \cite{childs2021theory,yi2022spectral}, this error ratio serves only as a relative indicator.

\end{document}